# Multi class intracellular protein targeting predictions in diatoms and other algae with complex plastids: ASAFind 2.0

Ansgar Gruber[1,2,*], Cedar McKay[3], Miroslav Oborník[1,2], Gabrielle Rocap[3]

[1]Biology Centre, Institute of Parasitology, Czech Academy of Sciences, Czech Republic

[2]Faculty of Science, University of South Bohemia, Czech Republic

[3]School of Oceanography, University of Washington, United States of America

[*]Correspondence: ansgar.gruber@paru.cas.cz

**Running Title**

Protein targeting predictions in diatoms

**Abstract**

Cells of diatoms and related algae with complex plastids of red algal origin are highly compartmentalized. These plastids are surrounded by four envelope membranes, which also define the periplastidic compartment (PPC), the space between the second and third membranes. The PPC corresponds to the cytosol of the eukaryotic alga that was the ancestor of the complex plastid. Metabolic reactions as well as cell biological processes take place in this compartment; however, its exact function remains elusive. Automated predictions of protein locations proved useful for genome wide explorations of metabolism in the case of plastid proteins, but until now, no automated method for the prediction of PPC proteins was available. Here, we present an updated version of the plastid protein predictor ASAFind, which includes optional prediction of PPC proteins. The new ASAFind version also accepts the output of the most recent versions of SignalP (5.0) and TargetP (2.0) input data. Furthermore, we release a Python script to calculate custom scoring matrices for adjustment of the ASAFind method to other groups of algae, and included the option to run the predictions with custom scoring matrices in a simplified score cut-off mode.

**Introduction**

Eukaryotic cells are characterized by the presence of membrane bound organelles, which fulfil distinct functions in the cell. The location of a protein in the cell has important consequences for its function. Therefore, *in silico* prediction software for intracellular protein locations became important in the analyses of biological sequence data, on the (meta-) genome or (meta-) transcriptome levels (Emanuelsson, et al. 2007; Nielsen, et al. 2019).

In photosynthetic eukaryotes, photosynthesis as well as other important metabolic pathways take place in the plastids (chloroplasts) (Gould, et al. 2008). These organelles are of endosymbiotic origin, they evolved from cyanobacteria, which were taken up by a eukaryotic cell, and transformed into organelles. However, in the majority of photosynthetic organisms, plastid evolution was more convoluted and involved one or several eukaryote – eukaryote endosymbioses (Archibald 2009, 2015; Oborník 2019). This led to a variety of cellular topologies that can be found in the resulting organisms with these, so-called, 'complex' plastids (Kroth 2002; Gould, et al. 2008). Also the mechanisms and signals for protein targeting differ between cells with complex plastids, and cells that contain primary plastids that directly trace back to cyanobacteria (Kroth 2002; Patron and Waller 2007; Gould, et al. 2008; Novák Vanclová, et al. 2020). Complex plastids are characterized by the presence of an additional cell compartment, the periplastidic compartment (PPC). The PPC corresponds to the former cytosol of the eukaryotic endosymbiont, which gave rise to the plastid. Most knowledge about the PPC comes from cryptophyte algae, which harbour a residual nucleus of the former endosymbiont, called the nucleomorph, inside the PPC (Gould, et al. 2008; Archibald 2009). From the genes encoded on the nucleomorph, it can be concluded that cell biological and metabolic processes take place in the PPC (Curtis, et al. 2012). In diatoms, which also contain complex plastids of red algal origin, the function of the PPC is more elusive (Gruber, et al. 2009; Weber, et al. 2009), although the PPC is probably an important player in the metabolic network of the cell (Tachibana, et al. 2011; Flori, et al. 2016; Moog 2018). Despite the fact that many PPC proteins in diatoms are known (Moog, et al. 2011), currently there is no method for the prediction of PPC proteins from genome data.

Primary plastids are found in green algae and higher plants, in red algae (rhodophytes), and in glaucophyte algae (Reyes-Prieto, et al. 2007; Oborník 2019). *In silico* targeting predictions for nucleus encoded proteins in these organisms are available, and have been used in genome annotations and functional genomics (Tardif, et al. 2012; Grossman, et al. 2019).

Organisms with rhodophyte-derived complex plastids are particularly diverse, and distributed worldwide in marine and freshwater habitats (Archibald 2009; de Vargas, et al. 2015). In this group of algae, which includes stramenopiles, haptophytes and cryptophytes, protein targeting to the plastids starts with passage of the pre-protein through the endoplasmic reticulum (ER). Nucleus encoded plastid proteins consequently are translated on membrane bound ribosomes, and possess cleavable signal peptides, that are exchangeable with ER signal peptides. The signal peptide is followed by a transit peptide, with similar features as transit peptides of proteins targeted to primary plastids, forming a bipartite targeting signal (Apt, et al. 2002; Kroth 2002; Patron and Waller 2007). For this reason, plastid protein prediction in organisms with complex plastids cannot be achieved with the prediction tools developed for plastid proteins of higher plants and algae with primary plastids (Gruber and Kroth 2014; Gruber and Kroth 2017; Moog 2018). Two dedicated plastid protein prediction tools have been developed for heterokont algae, Hectar (Gschloessl, et al. 2008) and ASAFind (Gruber, et al. 2015). ASAFind is an open source software which uses signal peptide predictions as input data, and employs a scoring matrix of proteins that are exclusively from diatoms (Gruber, et al. 2015). Modification of this scoring matrix to one that has been compiled from the target organism, increases the performance in this organism (Füssy, et al. 2019). Here, we present a new version of ASAFind, which (i), includes predictions of PPC targeted proteins, (ii) allows for an easy calculation of custom scoring matrices for target organisms with specific requirements and, (iii) accepts the results of newly published signal peptide predictors as input data.

**Results and Discussion**

ASAFind builds its prediction on the identification of a conserved sequence motif, which surrounds the signal peptide cleavage site (Kilian and Kroth 2005). The sequence conservation is partly due to the presence of a signal peptide cleavage site motif (see -1 and -3 S rule for signal peptides (von Heijne 1983)), and partly due to the presence of a phenylalanine (or similar) residue at the N-terminus of the transit peptide (Kilian and Kroth 2005; Gruber, et al. 2007), a sequence feature that is known from plastid transit peptides in red- and glaucophyte algae, and from most groups of algae with complex plastids of red algal origin (Patron and Waller 2007). ASAFind identifies the motif by collecting scores according to a matrix derived from diatom plastid proteins, while scanning the region surrounding the signal peptide cleavage site that has been predicted by the input data (Gruber, et al. 2015). ASAFind 1 accepts the output of SignalP 3 and SignalP 4 as input data. Recently, a new version of SignalP (SignalP 5 (Almagro Armenteros, Tsirigos, et al. 2019)) has been published, and also a new version of the multi-class predictor TargetP (TargetP 2) has been developed, that uses a similar approach as SignalP 5 and reaches similar prediction performance (Almagro Armenteros, Salvatore, et al. 2019). Combination of the new SignalP and TargetP versions with ASAFind was advantageous for the performance of ASAFind, and TargetP 2.0 also performs well in the prediction of mitochondrial proteins of diatoms (Gruber, et al. 2020).

For the new version of ASAFind, we therefore included the option of using SignalP 5 and TargetP 2.0 output for the plastid protein prediction (File S1).

ASAFind in the original version uses a scoring matrix that was calculated from diatom plastid proteins, and uses the information content of each position as well as the frequency of the residue found in this position (according to (Schneider and Stephens 1990)). ASAFind also proved useful for the prediction of plastid proteins in other groups of algae with complex plastids (Curtis, et al. 2012; Dorrell, et al. 2017; Füssy, et al. 2019; Wetherbee, et al. 2019), however, the prediction performance is lower than in diatoms (Curtis, et al. 2012; Füssy, et al. 2019), but in many cases remains unknown. In order to facilitate protein targeting predictions in other organisms than diatoms, we included the feature of using custom scoring matrices in the new ASAFind version 2.0 (File S1). This approach has been used by Füssy, et al. (2019)), who found that it improves the prediction of *Chromera velia* and *Vitrella brassicaformis* plastid proteins. We included a custom scoring matrix generator with the new ASAFind, that calculates a scoring matrix from a FASTA file of training proteins (File S2). In the new custom scoring matrix generator, also a small sample correction as suggested by Schneider, et al. (1986)) and Schneider and Stephens (1990)) is applied, and we updated the diatom plastid protein scoring matrix used by ASAFind as a standard accordingly (File S3). Because the score cut-off for the 'plastid, high confidence' category of ASAFind is calibrated according to the score on the default scoring matrix (Gruber, et al. 2015), the new ASAFind includes a simple score cut-off mode, in which no distinction between confidence levels is made. This simple score cut-off mode facilitates adjustment of ASAFind to other organisms and scoring matrices (File S1).

Protein targeting to the PPC, in the same way as plastid targeting, starts with ER import of the protein (Gould, Sommer, Kroth, et al. 2006). While the PPC is the destination for PPC targeted proteins, also plastid proteins pass through the PPC during targeting (Gould, et al. 2008). The targeting signals of PPC proteins are very similar to plastid targeting signals; in GFP fusion experiments, a single amino acid exchange resulted in re-targeting of a protein to the plastid (Gould, Sommer, Kroth, et al. 2006). Comparing the bipartite targeting pre-sequences of plastid and PPC proteins, the clearest difference seems to be the lack of a conserved phenylalanine (or similar) residue in the + 1 position of the transit peptide (Gould, Sommer, Kroth, et al. 2006; Gruber, et al. 2009; Weber, et al. 2009; Moog, et al. 2011). We therefore decided to include scoring of PPC transit peptides to ASAFind, and collected a set of PPC proteins that has no overlap to the reference set for testing the method (Figure 1, Table S4, see Methods for details). As expected, in this set of sequences, phenylalanine or similar residues are not observed in the +1 position of the transit peptide, while the signal peptide cleavage site motif upstream of the transit peptide is conserved in a similar way as in the plastid targeting pre-sequences (Figure 1). Furthermore, another difference between plastid and PPC targeting pre-sequences became apparent; in plastid proteins, the fist 15 positions of transit peptides are almost devoid of negatively charged residues, while in PPC proteins, negatively charge residues are present and evenly distributed across the whole 25 position sequence window (Figure 1). This is in line with the experimental finding that positive net-charges are required for protein transport to the plastid, but not to the PPC (Felsner, et al. 2010).

Optional scoring of sequences for PPC targeting was implemented into ASAFind 2, analogously to the collection of a 20 residue plastid transit peptide score starting from the position that gave the highest cleavage site score during the sliding window cleavage site motif identification (described in more detail in (Gruber, et al. 2015)). The complete method follows the decision tree shown in Figure 2. With this sequential procedure, PPC prediction profits from the high specificity and sensitivity of the preceding plastid protein prediction step.

In order to evaluate the performance of the plastid and PPC protein predictions, we updated the reference set of experimentally localized *Phaeodactylum tricornutum* proteins (Gruber, et al. 2015) with newly published experimental localisations of proteins (Table S6, detailed in Methods). The new reference set consists of 262 proteins, which could unambiguously be classified to the following categories: plastid, 71 sequences; PPC, 44 sequences; other signal peptide containing proteins, 92 sequences; mitochondrial proteins, 26 sequences; other (no targeting pre-sequence), 29 sequences. This set is not only much larger than the previous version used by Gruber, et al. (2015)) (132 sequences, 55 plastid, 77 non-plastid, positive:negative ratio for plastid proteins is 0.71), but also more balanced between the categories (more non-plastid and non-PPC proteins, positive:negative ratios are 0.37 for plastid proteins, 0.2 for PPC proteins, 0.11 for mitochondrial proteins, and 3.76 for proteins with signal peptides).

According to the classifications given in Figure 2, we calculated multiclass correlation coefficients ($R_K$), which were positive and above 0.5 in all cases (Table 1, calculated from Table S7; $R_K$ can range between -1 and 1, 0 is indicative of a random assignment, 1 represents a perfect prediction). The addition of the plastid category by ASAFind led to an increased overall $R_K$ compared to the three categories already predicted by TargetP (Table 1). However, this is not the case if PPC protein prediction is included, indicating that the PPC protein prediction is less accurate than the predictions for the other categories (Table 1). Comparing the per-category parameters precision and recall (or sensitivity) (Table 2), it becomes evident that the PPC protein predictions are highly precise (79 % of positive predictions are correct), but not very sensitive (only 25 % of the PPC proteins are detected). This provokes the question if PPC proteins are indeed a homologous group, or if multiple targeting signals lead to the assembly of a seemingly homologous set of proteins. Furthermore, also the presence of experimental false positives (not from prediction, but from experimental artefacts) could explain this finding. It is therefore interesting to see that most of the PPC proteins from the reference set were experimentally localized in GFP fusion experiments (Table S6), following observation of a phenotype in the transformed cell lines that has ben described as 'blob'-like structure (BLS). This BLS phenotype, has first been observed in cells expressing GFP fused to mutated plastid targeting pre-sequences (Kilian and Kroth 2005), and is indistinguishable from the phenotype observed in cells expressing GFP fused to native PPC targeting pre-sequences (Gould, Sommer, Kroth, et al. 2006; Gruber, et al. 2009).

Furthermore, a very similar phenotype of intracellular GFP distribution has been observed in *Thalassiosira pseudonana*, when pre-sequences of cell wall proteins were fused to the GFP reporter (Poulsen, et al. 2013). Also in our *P. tricornutum* reference dataset, conflicting experimental results between the BLS phenotype in GFP fusion experiments, and extracellular detection of the protein in a proteomics study was observed in two cases (Table S6). In addition to this, ambiguous/conflicting results between BLS and ER experimental locations also occurred two times (Table S6). Therefore, classifications of proteins as PPC proteins should generally be seen with caution, in the case of predictions, as well as in the case of experimental results. Given the high precision of our current PPC protein prediction, and the biological interest in the PPC, we decided to release the PPC protein prediction feature despite its low sensitivity. ASAFind 2 also accepts custom scoring matrices for PPC protein predictions, which might help to further improve PPC protein predictions, or to adjust the method to other groups of organisms than diatoms.

Taken together, ASAFind 2.0, with the forwarding of mitochondrial predictions from TargetP, and with the inclusion PPC protein predictions, allows for multi class protein targeting predictions in diatoms in two simple steps, the collection of TargetP results, and the subsequent run of ASAFind.

**Methods**

ASAFind 2.0 was developed from the original ASAFind code published in Gruber, et al. (2015)). Work on ASAFind 2 started with ASAFind.py version 1.1.7. The current version at the time of publication (2.0) is included in File S1, future updates will be made available via the online repository. During the updates, ASAFind.py was made fully Python 3 compatible.

For the calculation of custom scoring matrices, we included the Python script score_table.py (File S2), which is also available via the above-mentioned online repository.

For the calculation of a scoring matrix for PPC proteins, three sets of experimentally localized proteins were assembled, direct experimental results obtained in *Thalassiosira pseudonana* (Samukawa, et al. 2014; Tanaka, et al. 2014), or in *Phaeodactylum tricornutum* (the 35 unambiguous BLS phenotype sequences listed in Table S3 of Gruber, et al. (2015))), and heterologous expression experiments of *Guillardia theta* sequences tested in *P. tricornutum* (Gould, Sommer, Hadfi, et al. 2006; Gould, Sommer, Kroth, et al. 2006). *Guillardia theta* homologues of the *P. tricornutum* set were taken from Curtis, et al. (2012)). Other best BLAST hits of proteins from the experimental sets were identified in the non-redundant *G. theta* gene catalog (Curtis, et al. 2012), and in the optimized non redundant *T. pseudonana* gene catalog ("Dataset S1" of Gruber, et al. (2015)), developed from the genome data published by Armbrust, et al. (2004))). Only the best BLAST (Altschul, et al. 1997) hit of each sequence in each genome was selected, and only further considered if it could be reasonably considered a homologue, if prior knowledge about metabolism would be in line with the hit being a PPC protein, if the hit had a signal peptide, and if it was not predicted to be a plastid protein (see Table S5for details).

Sequence logos (Schneider and Stephens 1990) and frequency plots were calculated using the WebLogo online service (http://weblogo.berkeley.edu/, Crooks, et al. (2004))).

The reference set from Gruber, et al. (2015)), who compiled the results of 32 publications (Liaud, et al. 2000; Apt, et al. 2002; Domergue, et al. 2003; Kilian and Kroth 2004, 2005; Kroth, et al. 2005; Tanaka, et al. 2005; Gould, Sommer, Kroth, et al. 2006; Materna 2006; Gruber, et al. 2007; Lepetit, et al. 2007; Siaut, et al. 2007; Sommer, et al. 2007; Kitao, et al. 2008; Ast, et al. 2009; Burmeister 2009; Gruber, et al. 2009; Hempel, et al. 2009; Kitao and Matsuda 2009; Weber, et al. 2009; Bullmann, et al. 2010; Felsner, et al. 2010; Hempel, et al. 2010; Joshi-Deo, et al. 2010; Allen, et al. 2011; Bruckner, et al. 2011; Grouneva, et al. 2011; Moog, et al. 2011; Tachibana, et al. 2011; Vugrinec, et al. 2011; Allen, et al. 2012; Sturm, et al. 2013) to 134 experimentally verified protein localizations, was supplemented with localization data from 28 studies that have been published since then (Montsant, et al. 2005; Stork, et al. 2012; Peschke, et al. 2013; Schellenberger Costa, et al. 2013; Gile, et al. 2015; Lau, et al. 2015; Moog, et al. 2015; Buhmann, et al. 2016; Huang, et al. 2016; Lau, et al. 2016; Liu, et al. 2016; Niu, et al. 2016; Balamurugan, et al. 2017; Schreiber, et al. 2017; Chen, et al. 2018; Ewe, et al. 2018; Hao, et al. 2018; Huang, et al. 2018; Mix, et al. 2018; Rio Bartulos, et al. 2018; Wang, et al. 2018; Erdene-Ochir, et al. 2019; Shao, et al. 2019; Dell'Aquila, et al. 2020; Jallet, et al. 2020; Leyland, et al. 2020; Marter, et al. 2020; Seo, et al. 2020), resulting in a total of 284 proteins with experimentally verified intracellular locations (Table S6, without lipid droplet proteins). Protein sequences were added according to the *P. tricornutum* optimized gene catalog ("Dataset S2" of Gruber, et al. (2015))), developed from the genome data published by Bowler, et al. (2008))). Proteins localized via isolation of lipid droplets were not considered for statistical calculations for two reasons; one is that proteins can be targeted to lipid droplets via the endomembrane system or via the cytosol (Leyland, et al. 2020), and hence cannot be categorically classified as signal peptide containing or not, the other reason is that among the 662 proteins identified by Leyland et al. (2020) (without excluding proteins based on pre-sequence prediction, "jpy13063-sup-0008-TableS3.xls" from Leyland, et al. (2020))), 37 proteins were independently localized in other experimental studies. None of these proteins were confirmed as lipid droplet protein in the respective other study, and the conflicting results point to a high number of other intracellular destinations (Table S6). This indicates that the sensitivity of the method applied by Leyland et al. (2020) was high enough to also identify contaminating proteins from other intracellular locations. Localization data was only used for statistical analyses if the protein could be unambiguously assigned to one of the result categories (categories in Fig. 2, see Table S6 for details on the protein assignments). Proteins with identical N-termini (first 60 residues) were combined to just one sequence in the reference set. Table S6 contains information on all proteins, including the ones that were not considered for the statistical analyses; Table S7 contains only the final reference set (262 sequences). TargetP 2.0 (Almagro Armenteros, Salvatore, et al. 2019) was used in a portable version, which can be obtained via the TargetP 2.0 web site (https://services.healthtech.dtu.dk/service.php?TargetP). TargetP was used with the "non-plant" organism group selected, which is generally recommended for organisms with complex plastids, because otherwise, TargetP will also search for chloroplast transit peptides known from primary plastids of higher plants and green algae, a type of targeting pre-sequence that does not exist in organisms with complex plastids (Gruber and Kroth 2014; Gruber and Kroth 2017; Moog 2018; Gruber, et al. 2020).

For comparison, prediction results of SignalP 3.0 NN (Bendtsen, et al. 2004), SignalP 4.1 (Petersen, et al. 2011; Nielsen 2017) (with default and "sensitive" settings) and SignalP 5.0 (Almagro Armenteros, Tsirigos, et al. 2019) (with the "Eukarya" organism group selected) were obtained via the SignalP web server (http://www.cbs.dtu.dk/services/SignalP/).

Hectar (Gschloessl, et al. 2008) predictions were obtained from the Galaxy (Goecks, et al. 2010) web service provided by the Station Biologique de Roscoff in collaboration with ABiMS (https://webtools.sb-roscoff.fr/). Hectar has five possible result categories: "chloroplast", "signal peptide", "signal anchor", "mitochondrion", and "other localisation". Because our dataset contains no experimentally confirmed type II signal anchor protein, and since all proteins with this prediction result are in compartments that are reached via the secretory pathway, we combined the categories "signal peptide" and "signal anchor" to "secretory pathway", and counted the predictions as correct if the experimental location was in a compartment of the secretory pathway other than the plastid, see Table S8 for classifications.

Prediction statistics were calculated with help of the Python module scikit-learn (Pedregosa, et al. 2011), which follows the formulas given in (Matthews 1975; Baldi, et al. 2000; Gorodkin 2004).

## Author Contributions

Conceptualization, A.G., G.R. and M.O.; Methodology, A.G. and G.R.; Software, C.M., G.R. and A.G.; Investigation, A.G., C.M. and G.R.; Data Curation, A.G.; Validation, A.G.; Visualization, A.G.; Writing – Original Draft, A.G. Writing – Review & Editing, A.G., G.R., C.M. and M.O.

## Acknowledgements

We thank Zoltán Füssy for helpful discussions. This research was supported by the Institute of Parasitology of the Czech Academy of Sciences, by the Grant Agency of the Czech Republic (21-26115S), by the ERDF/ESF Centre for research of pathogenicity and virulence of parasites (No.CZ.02.1.01/0.0/0.0/16_019/0000759) and by the US National Science Foundation (G.R and C.M). Part of this work was carried out with the support of ELIXIR CZ Research Infrastructure (ID LM2018131, MEYS CR). The authors declare no conflict of interest.

# Figures

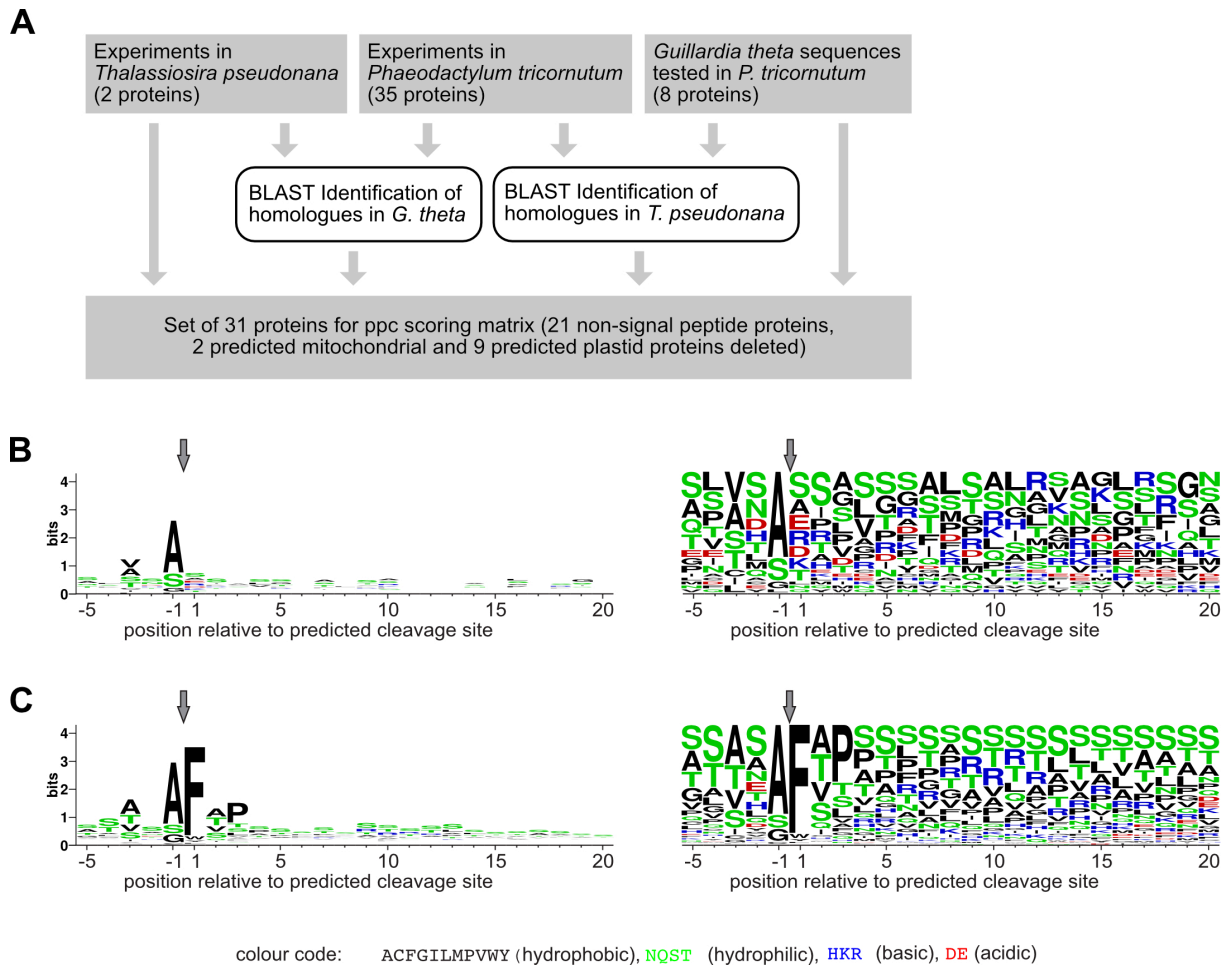

**Figure 1:** Scoring matrix for PPC protein prediction. **A** Composition of the sequence set; **B** sequence logo (left) and frequency plot (right) of the 31 sequence set used to calculate the PPC scoring matrix; **C** plastid protein scoring matrix (166 sequences, reproduced from Gruber, et al. (2015))).

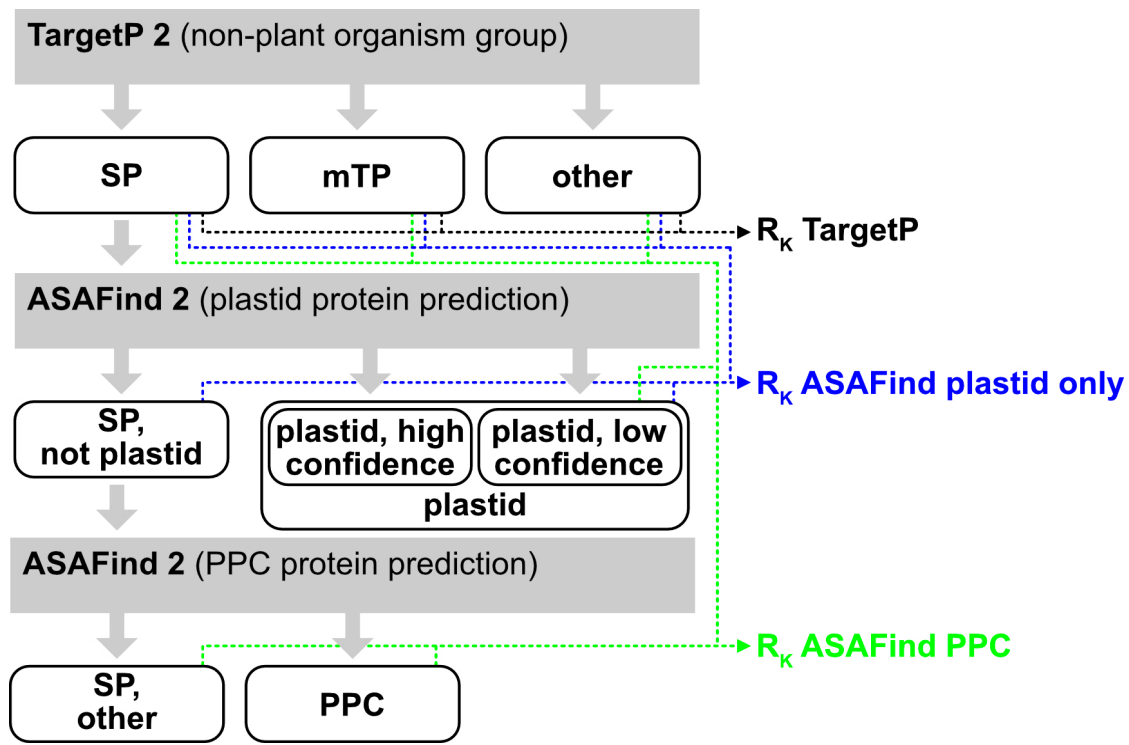

**Figure 2:** Decision tree of combinations of TargetP and ASAFind and corresponding correlation coefficients ($R_K$, see Table 1).

## Tables

**Table 1:** Correlation coefficients ($R_K$, according to Gorodkin (2004))) of multi-class targeting prediction methods (see Figure 2), tested with the reference set of experimentally localized *Phaeodactylum tricornutum* proteins (262 sequences, Table S6).

| Method | $R_K$ | Number of categories |
|---|---|---|
| ASAFind plastid only (with TargetP) | 0.61 | 4 |
| ASAFind PPC (with TargetP) | 0.52 | 5 |
| TargetP 2.0[1] | 0.57 | 3 |
| Hectar[2] | 0.59 | 4 |

[1] non-plant organism group

[2] "signal peptide" and "signal anchor" categories combined, see Methods for details

**Table 2:** Per-class prediction performance for all classes that can be predicted by ASAFind 2.0 in conjunction with TargetP 2.0, and in comparison to Hectar; tested with the set of experimentally localized *Phaeodactylum tricornutum* proteins (262 proteins, Table S6). Precision (positive predictive value), recall (sensitivity) and $F_1$ scores were calculated according to the given formulas, support (positives) indicates the number of proteins from each category in the reference set. TP: true positives, FP: false positives, FN: false negatives, P: positives, SP: signal peptide, PPC: periplastidic compartment, mTP: mitochondrial transit peptide.

| Category | Method | Precision $\frac{TP}{TP + FP}$ | Recall $\frac{TP}{TP + FN}$ | $F_1$ Score $\frac{2\, precision \times recall}{precision + recall}$ | Support P |
|---|---|---|---|---|---|
| Plastid, low or high confidence | TargetP/ ASAFind | 0.77 | 0.89 | 0.82 | 71 |
| Plastid, high confidence only | TargetP/ ASAFind | 0.91 | 0.72 | 0.80 | 71 |
| SP, not plastid | TargetP/ ASAFind | 0.94 | 0.50 | 0.65 | 136 |
| SP, not plastid, not PPC | TargetP/ ASAFind | 0.64 | 0.40 | 0.49 | 92 |
| PPC | TargetP/ ASAFind | 0.79 | 0.25 | 0.38 | 44 |
| SP | TargetP | 0.98 | 0.73 | 0.84 | 207 |
| mTP | TargetP | 0.80 | 0.77 | 0.78 | 26 |
| other | TargetP | 0.34 | 0.97 | 0.50 | 29 |
| Chloroplast | Hectar | 0.85 | 0.63 | 0.73 | 71 |
| Mitochondrion | Hectar | 0.82 | 0.69 | 0.75 | 26 |
| Secretory pathway[1] | Hectar | 0.78 | 0.72 | 0.75 | 136 |
| Other | Hectar | 0.46 | 0.97 | 0.62 | 29 |

[1] "signal peptide" and "signal anchor" categories combined, see Methods for details

**Supplemental Information**

**File S1:** ASAFind.py. Source code of ASAFind 2.0.

**File S2:** score_table.py. Script to calculate custom scoring matrices from FASTA files, input data needs to be trimmed to the intended sequence windows.

**Table S3:** Scoring matrix for plastid proteins, re-calculated from the sequence set used by Gruber, et al. (2015)), with small sample correction applied.

**Table S4:** Scoring matrix for PPC protein prediction, details on the selection of proteins. For experimental evidence in *P. tricornutum* proteins compare to S6, for other literature, digital object identifiers (dois) are given.

**Table S5:** Scoring matrix for PPC protein prediction, calculated from S4 via S2, with small sample correction applied.

**Table S6:** Reference set of experimentally localized *P. tricornutum* proteins.

**Table S7:** Prediction results for reference set and expected values for classification of the results in the calculation of statistical parameters.